\documentclass[12pt]{svmult6}
\usepackage{psfig}
\usepackage{epsf}
\begin{document}

\begin{center}
{\bf Non--parametric Inference in Astrophysics}\\ Larry Wasserman, Chris Miller,
Bob Nichol, Chris Genovese, Woncheol Jang, Andy Connolly, Andrew Moore, Jeff
Schneider and the PICA group.
\footnote{See www.picagroup.org for latest
software, papers and memberships of the PICA group.}
\end{center}

\begin{quote}
We discuss non--parametric density estimation
and regression
for astrophysics problems.
In particular,
we show how to compute non--parametric confidence intervals
for the location and size of peaks of a function.
We illustrate these ideas with recent
data on the Cosmic Microwave Background.
We also briefly discuss non--parametric Bayesian inference.
\end{quote}

\begin{center}
{\bf 1. Nonparametric Inference}
\end{center}

The explosion of data in astrophysics 
provides unique opportunities and challenges.
The challenges are mainly in data storage and manipulation.
The opportunities arise from the fact that
large sample sizes make nonparametric statistical methods 
very effective.
Nonparametric methods 
are statistical techniques that make 
as few assumptions as possible
about the process that generated the data.
Such methods are inherently more flexible
than more traditional parametric methods
that impose rigid and often unrealistic assumptions.
With large sample sizes, nonparametric methods
make it possible to find subtle effects
which might otherwise be
obscured by the assumptions built into parametric methods.
We begin by discussing two prototypical astrostatistics problems.

\vspace{.5cm}

{\sc Problem 1. Density Estimation.}
Let $X_1, \ldots, X_n$ denote the positions of $n$
galaxies in a galaxy survey.
Let $f(x)dx$ denote the probability of finding
a galaxy in a small volume around $x$.
The function $f$ is a {\em probability density function}, satisfying
$f(x) \geq 0$ and $\int f(x) dx=1$.
We regard
$X_1, \ldots, X_n$ as $n$ random draws from $f$.
Our goal is to estimate $f(x)$ from the data
$(X_1, \ldots, X_n)$ while making
as few assumptions about $f$ as possible.
Figure 1 shows redshifts from a 
pencil beam from the Sloan Digital Sky Survey.
The figure shows several nonparametric density estimates 
that will be described in more detail in Section 3.
The structure in the data is evident
only if we smooth the data by just the 
right amount (lower left plot).\footnote{
The data involve selection bias since
we can only observe brighter objects for larger redshifts.
However, the sampling is fairly complete out to about $z=0.2$.}

{\sc Problem 2. Regression.}
Figures 2 and 3
show cosmic microwave background (CMB) data
from
BOOMERaNG (Netterfield et al. 2001),
Maxima (Lee et al. 2001) and 
DASI (Halverson 2001).
The data consist of $n$ pairs
$(X_1, Y_1), \ldots, (X_n,Y_n)$.
Here,
$X_i$ is multipole moment
and $Y_i$ is the estimated power spectrum
of the temperature fluctuations.
If $f(x)$ denotes the true power spectrum
then
$$
Y_i = f(X_i) + \epsilon_i
$$
where $\epsilon_i$ is a random error with mean 0.
This is the standard {\em regression model.}
We call $Y$ the {\em response} variable and $X$ the
{\em covariate}.
Other commonly used names for $X$ include
{\em predictor} and {\em independent} variable.
The function $f$ is called the {\em regression function}.
The goal
in nonparametric regression is to estimate $f$
making only minimal smoothness assumptions about $f$.

The main messages of this paper are:
(1) with large data sets
one can estimate a function $f$
{\em nonparametrically}, that is, without assuming
that $f$ follows some given functional form;
(2) one can use the data
to estimate the optimal amount of smoothing;
(3) one can derive confidence sets for $f$
as well as confidence sets for interesting features of $f$.
The latter point is very important and is an example
of where rigorous statistical methods
are a necessity;
the usual confidence intervals
of the form
``estimate plus or minus error'' will not suffice.

\vspace{.5cm}

The outline of this paper is as follows.
Section 2 discusses 
some conceptual issues.
Section 3 discusses kernel density estimation. 
Section 4 discusses nonparametric regression.
Section 5 explains something that might be less familiar to astrophysicists,
namely, nonparametric estimation via shrinkage.
Section 6 discusses nonparametric confidence intervals.
In Section 7 we briefly discuss nonparametric 
Bayesian inference.
We make some concluding remarks in Section 8.

\vspace{.5cm}

{\bf Notation:}
We denote the mean of a random quantity $X$ by
$E(X)$, often written as $\langle X\rangle$ in physics.
The variance of $X$ is denoted by
$\sigma^2 \equiv Var(X) = E( X - E(X))^2$.
A random variable $X$ has a Normal (or Gaussian)
distribution with mean $\mu$ and variance $\sigma^2$, 
denoted by $X\sim N(\mu,\sigma^2)$,
if
$$
Pr(a < X < b) = \int _a^b \frac{1}{\sigma \sqrt{2\pi}}
\exp\left\{ -\frac{1}{2\sigma^2} (x-\mu)^2 \right\} dx.
$$
We use $\hat{f}$ to denote
an estimate of a function $f$.

\begin{center}
{\bf 2. Some Conceptual Issues}
\end{center}

{\sc 2.1. The Bias-Variance Tradeoff.}
In any nonparametric problem, we need to find methods that
produce estimates $\hat{f}$ of the unknown function $f$.
Obviously, we would like $\hat{f}$ to be close to $f$.
We will measure closeness with squared error:
$$
L(f,\hat{f}) = \int ( f(x) - \hat{f}(x))^2 dx.
$$
The average value of the error is called the {\em risk}
or {\em mean squared error} (MSE) and is denoted by:
$$
R(f,\hat{f}) = E_f \left[ L(f,\hat{f}) \right].
$$
A simple calculation shows that 
$$
R(f,\hat{f}) = \int {\rm Bias}^2_x \ dx + \int {\rm Var}_x \ dx
$$
where ${\rm Bias}_x = E[ \hat{f}(x)] - f(x)$
is the bias of $\hat{f}(x)$ and
${\rm Var}_x = Var[\hat{f}(x)] =
E[ (\hat{f}(x) - E[ \hat{f}(x)] )^2]$
is the variance of $\hat{f}(x)$.
In words:
$$
{\rm RISK} = {\rm BIAS}^2 + {\rm VARIANCE}.
$$

Every nonparametric method involves some sort of data-smoothing.
The difficult task in nonparametric inference
is to determine how much smoothing to do.
When the data are over-smoothed, the bias term is large and
the variance is small.
When the data are under-smoothed
the opposite is true; see Figure 4.
This is called the {\em bias-variance tradeoff}.
Minimizing risk corresponds to
balancing bias and variance.

\vspace{.5cm}

{\sc 2.2. Nonparametric Confidence Sets}.
Let $f$ be the function of interest, for example,
the true power spectrum in the CMB example.
Assume that $f\in {\cal F}$
where ${\cal F}$ is some very large class of functions.
A valid (large sample) $1-\alpha$ confidence set $C_n$
is a set $C_n \subset {\cal F}$
such that
$$
\liminf_{n\rightarrow\infty} \inf_{f \in {\cal F}}
Pr (f \in C_n)\geq 1-\alpha
$$
where $n$ is sample size.
In words, $C_n$ traps the true function $f$
with probability approximately $1-\alpha$ (or greater).
In parametric models,
confidence intervals take the form
$\hat{\theta} \pm 2 \,{\rm se}$
where $\hat{\theta}$ is an estimate of a parameter
$\theta$ and se is the standard error of the estimate
$\hat{\theta}$.
Bayesian interval estimates take essentially the same form.
Nonparametric confidence sets are derived
in a different way as we shall explain later in the paper.

If prior information is available on $f$
then it can be included by restricting $C_n$.
For example,
if it is thought that $f$ has at most three peaks and two dips,
we replace $C_n$ with 
$C_n \cap {\cal I}$ where
${\cal I}$ is the set of functions with no more
than three peaks and two dips.

Having constructed the confidence set
we are then in a position to give confidence intervals
for features of interest.
We express features as functions of $f$, written $T(f)$.
For example, $T(f)$ might denote the location
of the first peak in $f$.
Then
$$
\left( \inf_{f \in C_n} T(f),\sup_{f \in C_n} T(f)\right)
$$
is a $1-\alpha$ confidence interval for the feature $T(f)$.
In fact, we can construct valid, simultaneous
confidence intervals for many features of interest this way,
once we have $C_n$.
In section 6, we report such intervals
for the CMB data.

Let us dispel a common criticism about confidence
intervals.
An oft cited but useless interpretation of a 95 per cent
confidence interval is:
if we repeated the experiment many times,
the interval would contain the true value
95 per cent of the time.
This interpretation leads many researchers to
find confidence sets to be irrelevant since 
the repetitions are hypothetical.
The correct interpretation is:
if the method for constructing $C_n$ is used
on a stream of (unrelated) scientific problems,
we will trap the true value 95 per cent of the time.
The latter interpretation is correct
and is more scientifically useful than the former.

\vspace{.5cm}

{\sc 2.3. Where is the Likelihood?}
The likelihood function, which is a familiar centerpiece
of statistical inference in parametric problems,
is notably absent in most nonparametric methods.
It is possible to define a likelihood and even 
perform Bayesian inference in nonparametric problems.
But for the most part, likelihood and Bayesian methods
have serious drawbacks in nonparametric settings.
See section 7 for more discussion on this point.

\begin{center}
{\bf 3. Kernel Density Estimation.}
\end{center}

We now turn to problem 1, density estimation.
Let us start this section with its conclusion:
the choice of kernel (smoothing filter) is relatively
unimportant; the choice of bandwidth (smoothing parameter) is crucial;
the optimal bandwidth can be estimated from the data.
Let us now explain what this means.

Let $X_1, \ldots, X_n$ denote the observed data, a sample from $f$.
The most commonly used
density estimator is
the {\em kernel density estimator}
defined by
$$
\hat{f}(x) = \frac{1}{n}\sum_{i=1}^n \frac{1}{h}K \left( \frac{x-X_i}{h}\right)
$$
where $K$ is called the {\em kernel} and
$h$ is called the {\em bandwidth}.
This amounts to placing a smoothed out lump of mass of size $1/n$
over each data point $X_i$.
Excellent references on kernel density estimation
include Silverman (1986) and Scott (1992).

The kernel is usually assumed to 
be a smooth function satisfying
$K(x) \geq 0$,
$\int x K(x) dx =0$ and
$\tau \equiv \int x^2 K(x) dx >0$.
A fact that is well known in statistics
but appears to be less known in astrophysics is that
the choice of kernel $K$ is not crucial.
The optimal kernel that minimizes risk
(for large samples) is called the Epanechnikov kernel
$K(x) = .75 (1- x^2/5)/\sqrt{5}$ for $|x|< \sqrt{5}$.
But the estimates using another other smooth kernel
are usually numerically indistinguishable.
This observation is confirmed by theoretical calculations
which show that the risk is very insensitive to the
choice of kernel.
In this paper we use the Gaussian kernel
$K(x) = (2\pi)^{-1/2} e^{-x^2/2}$.

What does matter is the choice of bandwidth $h$
which controls the amount of smoothing.
Figure 1 shows the density estimate
with four different bandwidths.
Here we see how sensitive the estimate $\hat{f}$ is
to the choice of $h$.
Small bandwidths give very rough estimates while
larger bandwidths give smoother estimates.
Statistical theory tells us that, 
in one dimensional problems, 
\begin{eqnarray*}
R(f,\hat{f}) &=& {\rm BIAS}^2 \ + \ {\rm VARIANCE}\\
& \approx & \frac{1}{4}h^4 c_1 A(f) + \frac{c_2}{nh}
\end{eqnarray*}
where
$c_1 = \int x^2 K(x) dx$,
$c_2 = \int K(x)^2 dx$ and
$A(f) = \int (f^{\prime\prime}(x))^2 dx$.
The risk
is minimized by
taking the bandwidth equal to
$$
h_*= c_1^{-2/5} c_2^{1/5} A(f)^{-1/5} n^{-1/5}.
$$
This is informative because it tells us 
that the best bandwidth decreases at rate
$n^{-1/5}$ and leads to risk of order $O(n^{-4/5})$.
Generally, one cannot find a nonparametric estimator that
converges faster than $O(n^{-4/5})$.
This rate is
close to the rate of parametric estimators, namely,
$O(n^{-1})$.
The difference between these rates is the price we pay for being nonparametric.

The expression for $h_*$ depends on the unknown density $f$ which makes the result
of little practical use.
We need a data-based method for choosing $h$.
The most common method for
choosing a bandwidth $h$ from the data is
{\em cross-validation}.
The idea is as follows.

We would like to choose $h$ to minimize
the squared error
$\int ( f(x) - \hat{f}(x))^2 dz =
\int \hat{f}^2(x) dz -2 \int \hat{f}(x) f(x)dx + \int f^2(x)dx$.
Since
$\int f^2(x)dx$ does not depend on $h$,
this corresponds to minimizing
$$
J(h) = \int \hat{f}^2(x) dz -2 \int \hat{f}(x) f(x)dx.
$$
It can be shown that
$$
\hat{J}(h) = \int \hat{f}^2(x) dz -2 \frac{1}{n}\sum_{i=1}^n \hat{f}_{-i}(X_i).
$$
is an unbiased estimate of
$E[J(h)]$,
where
$\hat{f}_{-i}$ is the ``leave-one-out'' estimate obtained by omitting $X_i$.
Some algebra shows that
\begin{equation}\label{eq:cv}
\hat{J}(h) \approx
\frac{1}{h n^2} \sum_i \sum_j
K^*\left( \frac{ X_i-X_j}{h}\right) + \frac{2}{nh} K(0)
\end{equation}
where
$K^*(x) = K^{(2)}(x) - 2K(x)$
and
$K^{(2)}$ is the convolution of $K$ with itself.
Hence, it is not actually necessary to compute $\hat{f}_{-i}$.
We choose the bandwidth $\hat{h}$
that minimizes
$\hat{J}(h)$.
The lower left panel of figure 1 was based on cross-validation.
An important observation for large data bases is that
(\ref{eq:cv}) can be computed quickly using the fast Fourier transform;
see Silverman (1986, p 61-66).

\newpage

\begin{center}
{\bf 4. Nonparametric Kernel Regression}
\end{center}

Returning to the regression problem,
consider pairs of points
$(X_1,Y_1), \ldots, (X_n,Y_n)$ related by
$$
Y_i = f(X_i) + \epsilon_i.
$$
The kernel method for density estimation also
works for regression.
The estimate $\hat{f}$ is a weighted average of the
points near $x$: 
$\hat{f}(x) = \sum_{i=1}^n w_i Y_i$
where the weights are given by
$w_i \propto  K\left( \frac{ x-X_i}{h} \right)$.
This estimator is called the Nadaraya-Watson estimator.
Figure 2 shows that estimator for the CMB data.
Note the extreme dependence on the bandwidth $h$.

Once again, we use cross-validation to choose the bandwidth $h$.
The risk is estimated by
$$
\hat{J}(h) = \frac{1}{n} \sum_{i=1}^n (Y_i- \hat{f}_{-i}(X_i))^2.
$$

The first three panels in Figure 2 show the regression data
with different bandwidths.
The second plot is based on the cross-validation bandwidth.
The final plot shows 
the estimated risk 
$\hat{J}(h)$
from cross validation.
Figure 3 compares the nonparametric fit with the
fit by Wang, Tegmark and Zaldarriaga (2001).

Given the small sample size and the fact that we have
completely ignored the cosmological models
(as well as differential error on each data point)
the nonparametric fit does a remarkable job.
It ``confirms,'' nonparametrically, the existence
of three peaks, their approximate positions
and approximate heights.
Actually, the degree to which the fit confirms the three peaks
requires confidence statements that we discuss in section 6.

\begin{center}
{\bf 5. Smoothing by Shrinking}
\end{center}

There is another approach to nonparametric estimation 
based on
expanding $f$ into an
orthogonal series.
The idea is to 
estimate the coefficients of the series and then
``shrink'' these estimates towards 0.
The operation of shrinking is akin to smoothing.
These methods have certain advantages over
kernel smoothers.
First, the problem of estimating the bandwidth
is replaced with the problem of choosing the amount of shrinkage
which is, arguably,
supported by better statistical theory than the former.
Second, it is easier to construct valid confidence sets
for $f$ in this framework.
Third, in some problems one can choose the
basis in a well-informed way which will lead to
improved estimators.
For example, Donoho and Johnstone (1994, 1995)
and Johnstone (this volume)
show that wavelet bases can be used to great advantage
in certain problems.

Suppose 
we observe
$Y_i = f(x_i) + \epsilon_i$
where, for simplicity, we assume that
$x_1 =1/n, x_2 = 2/n , \ldots, x_n =1$.
Further suppose that
$\epsilon_i \sim N(0,\sigma^2)$.
Let $\phi_1, \phi_2, \ldots$ be an orthonormal basis 
for $[0,1]$:
$$
\int_0^1 \phi_j^2(x)dx =1\ \ \ {\rm and}\ \ \ 
\int_0^1 \phi_i(x) \phi_j(x)dx =0\ {\rm  when}\  i\neq j.
$$
For illustration, we consider
the cosine basis:
$\phi_1 (x) \equiv 1$,
$\phi_2 (x) = \sqrt{2} cos(\pi x)$,
$\phi_2 (x) = \sqrt{2} cos(2\pi x), \ldots$.
Expand $f$ in this basis:
$f(x) \sim \sum_{j=1}^\infty \beta_j \phi_j(x) \approx 
\sum_{j=1}^n \beta_j \phi_j(x).$
Estimating $f$ then amounts to estimating the $\beta_j$'s.
Let $Z_j = n^{-1/2}\sum_{i=1}^n Y_i \phi_j (i/n)$.
It can be shown that
$Z_j \approx N\left(\theta_j, \sigma^2\right), \ \ \ j = 1,\ldots, n$
where $\theta_j = \sqrt{n} \beta_j$.
Once we have estimates $\hat{\theta}_j$,
we set
$\hat{\beta}_j = n^{-1/2}\hat{\theta}_j$ and
$\hat{f}(x) = 
\sum_{j=1}^n \hat{\beta}_j \phi_j(x).$

How do we estimate $\theta = (\theta_1, \ldots, \theta_n)$ from
$Z = (Z_1, \ldots, Z_n)$?
A crude estimate is
$\hat{\theta}_j = Z_j$, $j=1, \ldots, n$.
This leads to a very noisy (unsmoothed) estimate of $f$.
Better estimates can be found by 
using {\em shrinkage} estimators.
The idea -- which goes back to James and Stein (1961) and Stein (1981) --
is to estimate $\theta$ by
shrinking the vector $Z$ closer to the origin.
A major discovery in mathematical statistics was that careful shrinkage
leads to estimates 
with much smaller risk.
Following Beran (2000) we consider 
shrinkage estimators of the form
$\hat{\theta} = (\alpha_1 Z_1, \alpha_2 Z_2, \ldots, \alpha_n Z_n)$
where $1\geq \alpha_1 \geq \alpha_2 \geq \cdots \geq \alpha_n \geq 0$
which forces more shrinkage 
for higher frequency cosine terms.

Let $\alpha = (\alpha_1, \ldots , \alpha_n)$ and
let $R(\alpha)$ denote the risk of $\hat{\theta}$ using shrinkage
vector $\alpha$.
An estimate of $R(\alpha)$,
called {\em Stein's unbiased risk estimate} (SURE), is
$$
\hat{R}(\alpha) =
\sum_j \left[\hat{\sigma}^2 \alpha_j^2 + (Z_j^2 - \hat{\sigma}^2)(1-\alpha_j)^2\right]
$$
where $\sigma^2$ has been estimated by
$\hat{\sigma}^2 = \frac{1}{k}\sum_{i=n-k+1}^n Z_i^2$
with $k < n$.
Using appropriate numerical techniques,
we minimize
$\hat{R}(\alpha)$ subject to the monotonicity constraint.
The minimizer is denoted by $\hat{\alpha}$ and the final estimate is
$\hat{\theta} = (\hat{\alpha}_1 Z_1, \hat{\alpha}_2 Z_2, \ldots, \hat{\alpha}_n Z_n)$.
Beran (2000) shows that the estimator 
obtained this way has
some important optimality properties.
Beran calls this approach
REACT (Risk Estimation, Adaptation, and Coordinate Transformation).
The estimated function $\hat{f}$ turns out to be similar to
the kernel estimator; due to space limitations we omit the plot.

\begin{center}
{\bf 6. Confidence Sets}
\end{center}

When estimating a scalar quantity $\theta$ with an estimator $\hat{\theta}$,
it is common to summarize the 
uncertainty for the estimate by reporting
$\hat{\theta} \pm 2 se$
where
${\rm se} \approx \sqrt{ Var(\hat{\theta})}$ is the
{\em standard error} of the estimator.
Under certain {\em regularity conditions},
this interval is a 95 per cent confidence interval, that is,
$$
Pr \left(\hat{\theta} - 2 \,{\rm se} \;\leq \;\theta \;\leq \;\hat{\theta} + 
2 \,{\rm se} \right)
\approx .95.
$$
This follows because, under the conditions alluded to above,
$\hat{\theta} \approx N(\theta, {\rm se}^2)$.

But the ``plus or minus 2 standard errors''
rule fails in nonparametric inference.
Consider estimating a density $f(x)$
at a single point $x$ with a kernel density estimator.
It turns out that
\begin{equation}\label{eq:dnorm}
\hat{f}(x) \approx N\left( f(x) + {\rm bias}, \frac{c_2 f(x)}{nh} \right)
\end{equation}
where
\begin{equation}\label{eq:bias}
{\rm bias} =\frac{1}{2} h^2 f^{\prime\prime}(x) c_1
\end{equation}
is the bias,
$c_1 = \int x^2 K(x) dx$ and
$c_2 = \int K^2(x) dx$.
The estimated standard error is 
\begin{equation}\label{eq:se}
{\rm se} = \left\{ \frac{ c_2 \hat{f}(x)}{nh} \right\}^{1/2}.
\end{equation}
Observe from (\ref{eq:dnorm}) that
$(\hat{f}(x) - f(x))/se \approx N( {\rm bias}/{\rm se},1)$.
If use the
``estimate plus/minus 2 se'' rule then
\begin{eqnarray*}
Pr\left( \hat{f}(x) - 2\,{\rm se} \leq f(x) \leq 
\hat{f}(x) + 2\,{\rm se} \right) & = &
Pr \left( -2 \leq
\frac{ \hat{f}(x) - f(x)}{ {\rm se} } \leq 2 \right)\\
& \approx &
Pr \left( -2 \leq N \left( \frac{{\rm bias}}{{\rm se}},1 \right) \leq 2\right).
\end{eqnarray*}
If ${\rm bias}/{\rm se}\rightarrow 0$ then this becomes
$Pr( -2 < N(0,1) < 2) \approx .95$.
As we explained in Section 2,
the optimal bandwidth is of the form
$h =c n^{-1/5}$.
If you plug 
$h =c n^{-1/5}$
this into (\ref{eq:bias}) and(\ref{eq:se})
you will see that
${\rm bias}/{\rm se}$ does not tend to 0.
The confidence interval
will have coverage less than .95.
In summary,
``estimate plus/minus 2 standard errors''
is not appropriate in nonparametric inference.
There are a variety of ways to
deal with this problem.
One is to use kernels with
a suboptimal bandwidth.
This undersmooths the estimate resulting
in a reduction of bias.

Another approach is based on the REACT
method (Beran and Dumbgen, 1998).
We construct 
a confidence set $C_n$
for the vector of function values
at the observed data,
${\bf f}_n = (f(X_1), \ldots, f(X_n))$.
The confidence set $C_n$ satisfies:
for any $c>0$,
$$
\limsup_{n\rightarrow \infty} \sup_{|| {\bf f}_n || \leq c   }
 | Pr(  {\bf f}_n \in C_n) - (1-\alpha)|\rightarrow 0
$$
where $||a|| = \sqrt{ n^{-1}\sum_i a_i^2 }$.
The supremum is important:
it means
that the accuracy of the coverage probability
does not depend on the true (unknown) function.

The confidence set, expressed in terms of the coefficients
$\theta$, is
$$
C_n =\left\{ \theta :\ n^{-1}\sum_j (\theta_j - \hat{\theta}_j)^2 \leq
\hat{R}_r + n^{-1/2}\hat{\tau}  z_\alpha \right\}
$$
where $z_\alpha$ is such that
$P(Z > z_\alpha)=\alpha$ where $Z\sim N(0,1)$
and $\hat{\tau}$ is a quantity computed from the data
whose formula we omit here.
Finally,
the confidence set for $f$ is
$$
{\cal D}_n = \left\{f:\  f = \sum_j \beta_j \phi_j :\ 
\beta_j = n^{-1/2}\theta_j , \ \theta \in {\cal C}_n \right\}.
$$

Let us return to the CMB example.
We constructed a 95 per cent confidence set $C_n$,
then we searched over $C_n$ and found the possible
number, location and heights of the peaks.
We restricted the search to functions with no more than 
three peaks and two dips as it was deemed unlikely that the
true power spectrum would have more than three peaks within the
range of scales presently covered by the balloon experiments.
Curves with just one or two peaks
cannot be ruled out at the 95 per cent level {\it i.e.} they are still
viable descriptions of the data but at a low statistical significance than
three peaked models.
The confidence intervals, restricted to three peak models, are as follows.

\begin{center}
\begin{tabular}{lll}
Peak & Location & Height\\ \hline
1    & (118,300) & (4361,8055) \\
2    & (377,650) & (1822,4798) \\
3    & (597,900) & (1839,4683) \\
\end{tabular}
\end{center}

The 95 per cent confidence interval for the
ratio of the height of the second peak divided
by the height of the first peak is
$(.21,1.4)$.
The 95 per cent confidence interval for the
ratio of the height of the third peak divided
by the height of the second peak is
$(.22,2.82)$.
Not surprisingly, the intervals are broad
because the data set is small. The reader is
referred to 
Miller et al (2002), for a more complete
discussion of this work and our final results {\it e.g.}
improvements in measurement error that are needed
to get more precise confidence sets.

\begin{center}
{\bf 7. Nonparametric Bayes}
\end{center}

There seems to be great interest in Bayesian methods
in astrophysics.
The reader might wonder if
it is possible to perform nonparametric Bayesian inference.
The answer is, sort of.

Consider estimating a density $f$
assumed to belong to
some large class
of functions such as
${\cal F} = 
\{ f : \ \int (f^{\prime\prime} (x))^2 dx \leq C \}$.
The ``parameter'' is the function $f$
and the likelihood function is
${\cal L}_n(f) = \prod_{i=1}^n f(X_i).$
Maximizing the likelihood leads to
the absurd density estimate that puts
infinite spikes on each data point.
It is possible to put a prior $\pi$ over ${\cal F}$.
The posterior distribution on ${\cal F}$ is
well defined and Bayes theorem still holds:
$$
Pr(f \in C \mid X_1, \ldots, X_n) =
\frac{ \int_C {\cal L}_n(f) d\pi (f)}
     { \int_{\cal F} {\cal L}_n(f) d\pi (f)}.
$$
Lest this seem somewhat abstract, take note
that much recent work in statistics
lately has led to methods for computing this
posterior.

However, there is a problem.
The parameter space ${\cal F}$
is infinite dimensional and, in such cases,
the prior $\pi$ is
extremely influential.
The result is that the posterior may concentrate
around the true function very slowly.
Worse, the 95 per cent Bayesian
credible sets will
contain the true function with very low frequency.
In many cases the frequency coverage probability 
of the Bayesian 95 per cent credible set is near 0!
Since high dimensional parametric models behave
like nonparametric models, these remarks
should give us pause before casually applying Bayesian methods
to parametric models with many parameters.

The results that make these
comments precise are fairly technical.
The interested reader is referred to
Diaconis and Freedman (1986),
Barron, Schervish and Wasserman (1999),
Ghosal, Ghosh and van der Vaart (2000),
Freedman (2000), Zhao (2000) and
Shen and Wasserman (2001).
The bottom line:
in nonparametric problems
Bayesian inference is an interesting research area
but is not (yet?) a practical tool.

\begin{center}
{\bf 8. Conclusion}
\end{center}

Nonparametric methods are at their best
when the sample size is large.
The amount and quality of astrophysics data
have increased dramatically in the last few years.
For this reason, we believe that
nonparametric methods will play an increasingly
important role in astrophysics.
We have tried to illustrate some of the key ideas
and methods here.
But we have really only touched on a few main points.
We hope through our continued interdisciplinary collaboration
and through others like it elsewhere, that
the development of nonparametric techniques
in astrophysics
will continue in the future.

\begin{center}
{\bf References}
\end{center}

\noindent
Barron, A., Schervish, M. and Wasserman, L. (1999).
The consistency of posterior distributions in nonparametric
problems. {\it The Annals of Statistics.} {\bf 27}, 536-561.

\noindent
Beran, R. (2000).
REACT Scatterplot Smoothers: Superefficiency Through Basis Economy.
{\em Journal of the American Statistical Association}.

\noindent
Beran, R. and Dumbgen, L. (1998).
Modulation Estimators and Confidence Sets.
{\em The Annals of Statistics}, 26, 1826-1856.

\noindent
Diaconis, P. and Freedman, D. (1986). On the
consistency of Bayes estimates (with discussion). {\it Ann. Statist.}
{\bf 14}, 1-67.

\noindent
Donoho, D.L. and Johnstone, I.M. (1994).
Ideal spatial adaptation via wavelet shrinkage.
{\em Biometrika}, 81, 425-455.

\noindent
Donoho, D.L. and Johnstone, I.M. (1995).
Adapting to unknown smoothness via wavelet shrinkage.
{\em Journal of the American Statistical Association}, 90, 1200-1224.

\noindent
Freedman, D. A. (1999).
On the Bernstein-Von Mises theorem with infinite dimensional parameters.
{\em The Annals of Statistics}, {\bf 27}, 1119-1140.

\noindent
Ghosal, S., Ghosh, J.K. and van der Vaart, A. (2000).
Convergence rates of posterior distributions.
{\em The Annals of Statistics}, {\bf 28}, 500-531.

\noindent
Halverson, N.W. et al. submitted to ApJ (2001) astro-ph/0104489

\noindent
James, W. and Stein, C. (1961).
Estimation with quadratic loss.
In: {\em Proc. Fourth Berkeley Symp. Math. Statist. Prob., 1, (J. Neyman, ed.)}
University of California Press, Berkeley. pp. 361--380. 

\noindent
Lee, A.T., et al. (2001) astro-ph/0104459

\noindent
Miller, C. J., Nichol, R. C., Genovese, C., Wasserman, L., 2002,
ApJ Letters, submitted (see astro--ph/0112049).

\noindent
Netterfield, C.B., et al. (2001)
astro-ph/0104460.

\noindent
Scott, D.W. (1992).
{\em Multivariate Density Estimation:
Theory, Practice, and Visualization}.
John Wiley, New York.

\noindent
Shen, X. and Wasserman, L. (2001).
Rates of convergence of posterior distributions.
To appear: {\em The Annals of Statistics}.

\noindent
Silverman, B.W. (1986).
{\em Density Estimation for Statistics and Data Analysis}.
Chapman-Hall: New York.

\noindent
Stein, C. (1981). 
Estimation of the mean of a multivariate normal distribution. 
{\em The Annals
of Statistics}, {\bf 9}, 1135-1151.

\noindent
Wang, X., Tegmark, M., and Zaldarriaga, M. (2001).
To appear in Physcial Review D. astro-ph/0105091.

\noindent
Zhao, L. H. (2000).
Bayesian aspects of some nonparametric priors.
{\em The Annals of Statistics}, {\bf 28}, 532-552.

\newpage

\begin{figure}
\psfig{figure=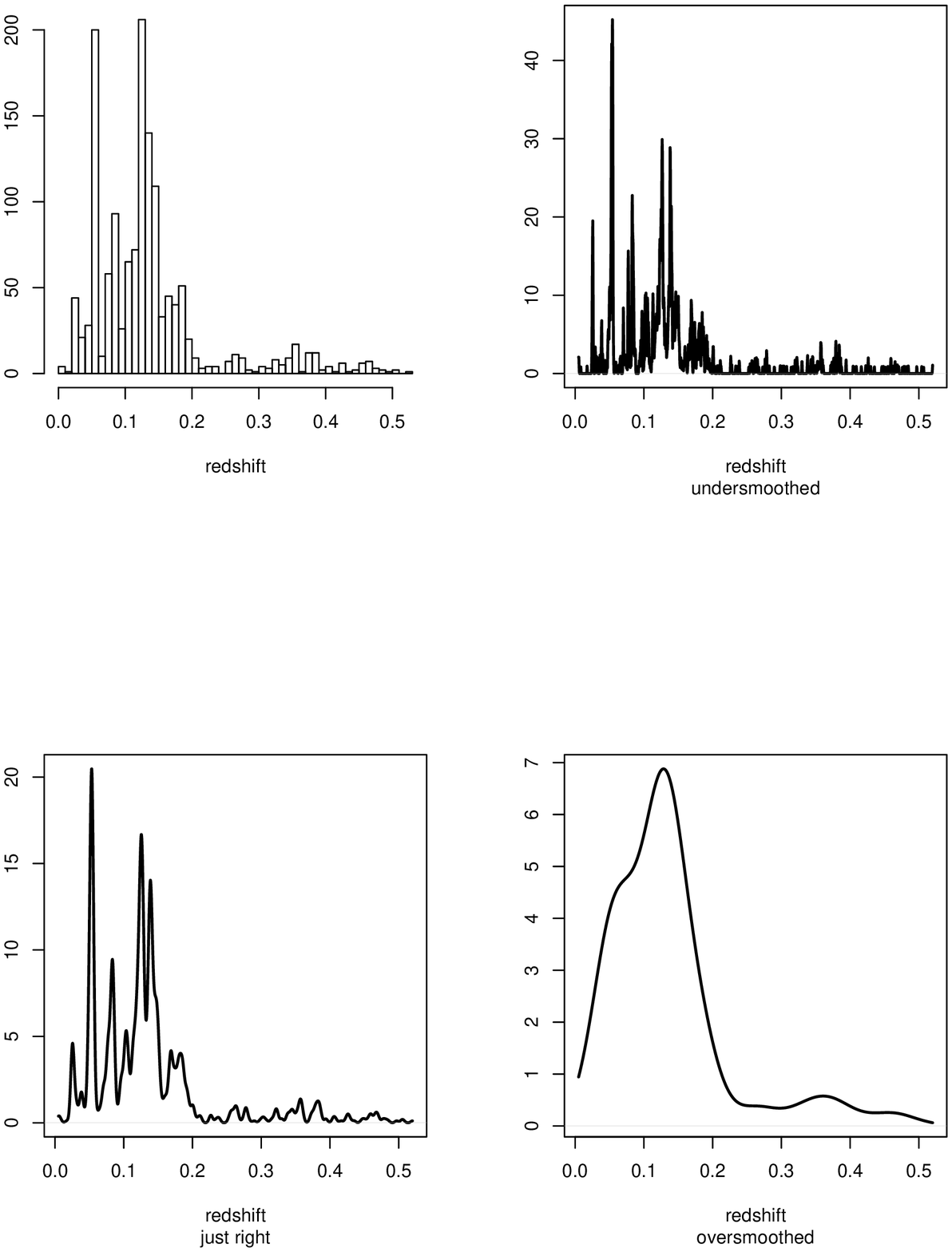,angle=-0,width=6in,height=6in}
\begin{quote}
Figure 1. Redshift data.
Histogram and three kernel density estimates based on three different
bandwidths. The bandwidth for the estimate in the lower left panel 
was estimated from the data using cross-validation.
\end{quote}
\end{figure}

\begin{figure}
\psfig{figure=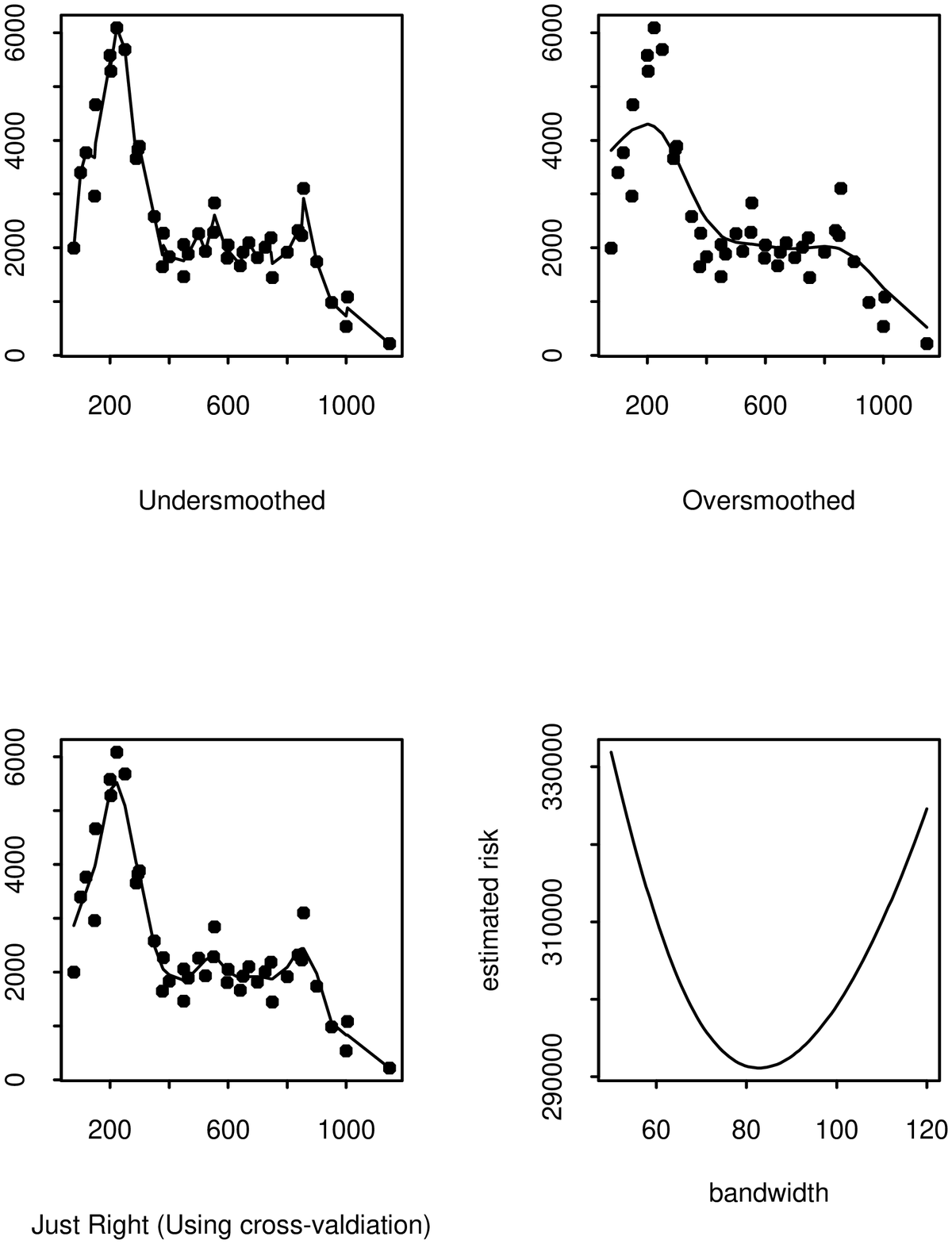,angle=-0,width=6in,height=6in}
\begin{quote}
Figure 2. CMB data. Section 4 explains the methods.
The first fit is undersmoothed, the second is oversmoothed and the third
is based on cross-validation.
The last panel shows the estimated risk versus the bandwidth of the smoother.
The data are from BOOMERANG, Maxima and DASI.
\end{quote}
\end{figure}

\begin{figure}
\psfig{figure=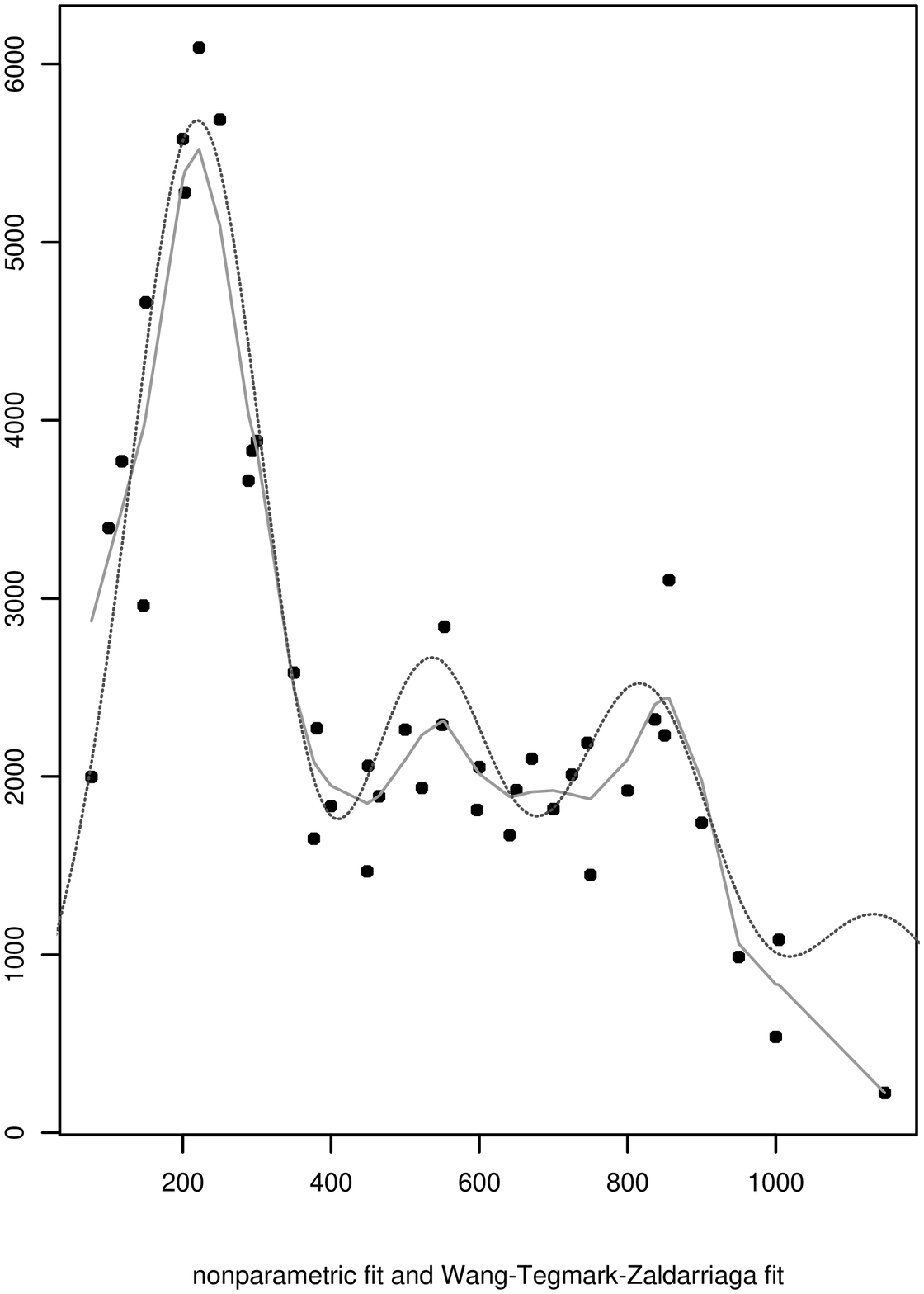,angle=-0,width=6in,height=6in}
\begin{quote}
Figure 3. 
Best nonparametric fit together with
parametric fit from
Wang, Tegmark and Zaldarriaga (2001). Please see Miller et al. (2002) for our
final, best--fit  results.
\end{quote}
\end{figure}

\begin{figure}
\psfig{figure=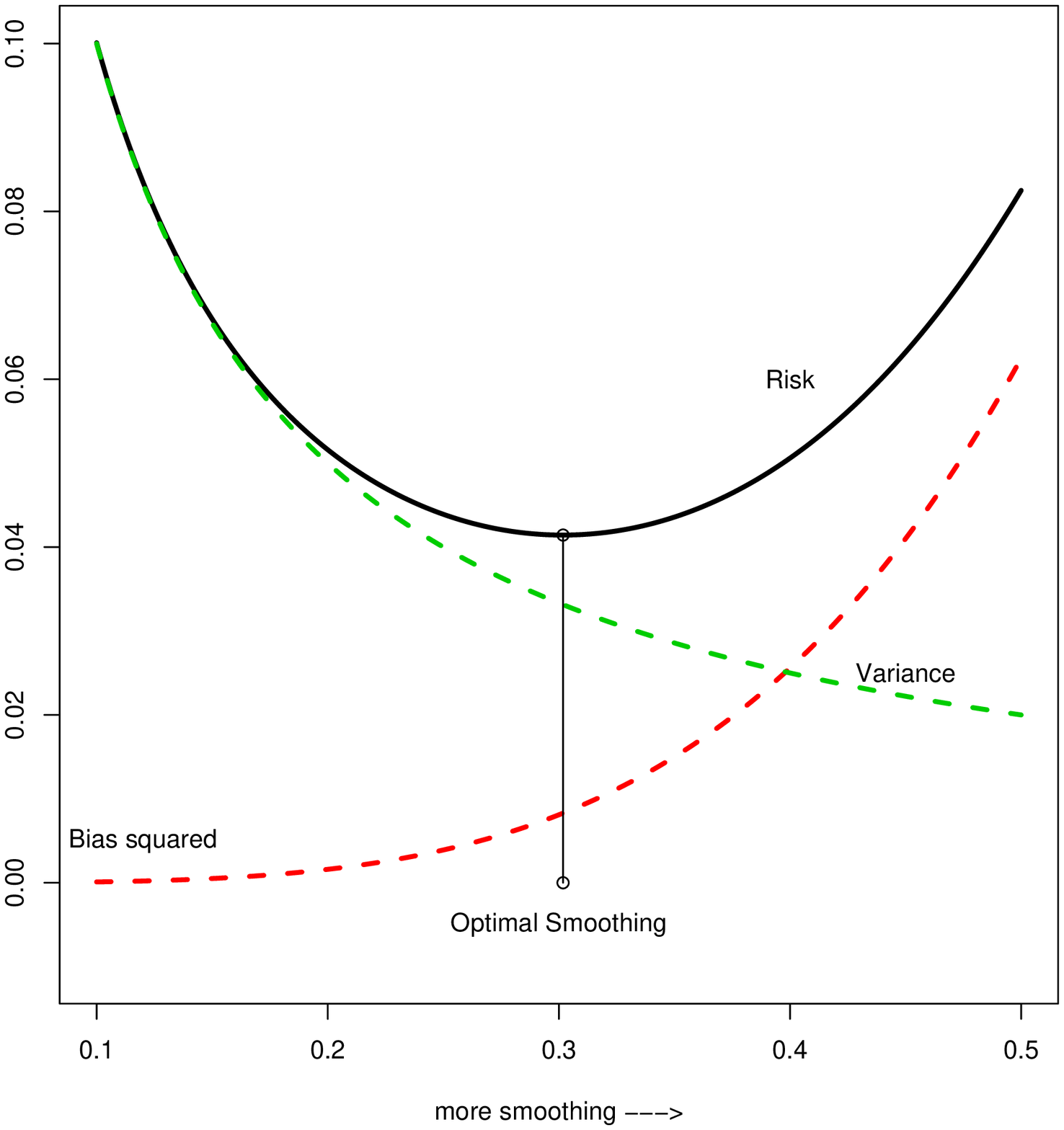,angle=-0,width=6in,height=6in}
\begin{quote}
Figure 4. The Bias-Variance tradeoff.
The bias increases and the variance decreases with the amount of smoothing.
The optimal amount of smoothing, indicated by the vertical line,
minimizes the risk = ${\rm bias}^2 + {\rm variance}$.
\end{quote}
\end{figure}

\end{document}